\documentclass[a4paper]{jpconf}
\usepackage{graphicx}
\usepackage{amssymb}
\usepackage{amsmath}
\usepackage{epsf}
\usepackage{bm}
\usepackage{cite}

\begin{document}

\title{Using Noether symmetries to specify f(R) gravity}
\author{Andronikos Paliathanasis}

\begin{abstract}
A detailed study of the modified gravity, f(R) models is performed, using
that the Noether point symmetries of these models are geometric symmetries
of the mini superspace of the theory. It is shown that the requirement that
the field equations admit Noether point symmetries selects definite models
in a self-consistent way. As an application in Cosmology we consider the
Friedman -Robertson-Walker spacetime and show that the only
cosmological model which is integrable via Noether point symmetries is the $%
\left( R^{b}-2\Lambda \right) ^{c}$ model, which generalizes the Lambda
Cosmology. Furthermore using the corresponding Noether integrals we compute
the analytic form of the main cosmological functions.\newline
Pacs - numbers:98.80.-k,95.35.+d,95.36.+x
\end{abstract}

\address{Faculty of Physics, Department of
Astronomy-Astrophysics-Mechanics,University of Athens, Panepistemiopolis, Athens 157 83, Greece}

\ead{anpaliat@phys.uoa.gr}

\section{Introduction}

The recent cosmological data indicate that the universe is spatially flat
and has suffered two acceleration phases. An early acceleration phase
(inflation), which occurred prior to the radiation dominated era and a
recently initiated accelerated expansion.

An easy way to explain this expansion is to consider an additional fluid
with negative equation of state parameter, usually called dark energy, that
dominates the universe at late times. In spite of that, the absence of a
fundamental physical theory, regarding the mechanism inducing the cosmic
acceleration, has given rise to a plethora of alternative cosmological
scenarios. Most of them are based either on the existence of new fields in
nature (dark energy) or in some modification of Einstein's general
relativity (GR), with the present accelerating stage appearing as a sort of
geometric effect (\textquotedblright geometrical\textquotedblright\ dark
energy).

The simplest dark energy probe is the cosmological constant $\Lambda $\
(vacuum) leading to the $\Lambda $CDM cosmology \cite%
{Weinberg89,Peebles03,Pad03}. However, it has been shown that $\Lambda $CDM
cosmology suffers from two major drawbacks known as \ the fine tuning
problem and the coincidence problem \cite{Peri08}. Besides $\Lambda $CDM
cosmology, \ many other candidates have been proposed in the literature,
such as time-varying $\Lambda (t)$\ cosmologies, quintessence, $k-$essence,
tachyons, modifications of gravity, Chaplygin gas and others \cite%
{Ratra88,Lambda4,Bas09c,Linder2004,LSS08,Brookfield2005td}.

There are other possibilities to explain the present accelerating stage. For
instance, one may consider that the dynamical effects attributed to dark
energy can be resembled by the effects of a nonstandard gravity theory. In
other words, the present accelerating stage of the universe can be driven
only by cold dark matter, under a modification of the nature of gravity.
Such a reduction of the so-called dark sector is naturally obtained in the $%
f(R)$ gravity theories \cite{Sot10}. In the original nonstandard gravity
models, one modifies the Einstein-Hilbert action with a general function $%
f(R)$ of the Ricci scalar $R$. The $f(R)$ approach is a relative simple but
still a fundamental tool used to explain the accelerated expansion of the
universe. A pioneering fundamental approach was proposed long ago with $%
f(R)=R+mR^{2}$\thinspace\ \cite{Star80}. Later on, the $f(R)$ models were
further explored from different points of view in \cite%
{Carrol,Amendola-2007a,Amendola-2007b} and indeed a large number of
functional forms of $f(R)$ gravity is currently available in the literature
\cite{Cap02,Noi,Hu07,Starobinsky-2007,Tsuj,AmeTsuj}.

The aim of the present work is to investigate which $f(R)$ models admit
extra Noether point symmetries and use the first integrals of these models
to determine analytic solutions of their field equations. The idea to use
Noether symmetries in cosmological studies is not new and indeed a lot of
attention has been paid in the literature (see \cite%
{Cap96,Cotsakis,RubanoSFQ,Sanyal05,Szy06,Cap07,Capa07,Bona07,Vakili08,TsamGRG,Basilakos11,Vakili11,Jamil11,Chirst12,Cap12,Wei2012}%
).

The main reasons for the consideration of this hypothesis is that (a) the
Noether point symmetries provide integrals, which assist the integrability
of the system, (b) is a geometric criterion because the Noether symmetries
associated with the geometry of the field equations. A fundamental approach
to derive the Noether point symmetries of a given dynamical system moving in
a Riemannian space has been proposed in \cite{Tsam10}. A similar analysis
can be found in \cite{Olver,StephaniB}.

The structure of the paper is as follows. The basic theoretical elements of
the problem are presented in section \ref{CMG}, where we also introduce the
basic FRW cosmological equations in the framework of $f(R)$ models. The
geometrical Noether point symmetries and their connections to the $f(R)$
models are discussed in sections \ref{LieN}. In section \ref{Analsol} we
provide analytical solutions for those $f(R)$ models which are Liouville
integrable via Noether point symmetries. In section \ref{Nonf} \ we study
the Noether symmetries in spatially non flat $f(R)$ cosmological models.
Finally, we draw our main conclusions in section \ref{Conc}.

\section{Cosmology with a modified gravity}

\label{CMG}

Consider the modified Einstein-Hilbert action:
\begin{equation}
S=\int d^{4}x\sqrt{-g}\left[ \frac{1}{2k^{2}}f\left( R\right) +\mathcal{L}%
_{m}\right]  \label{action1}
\end{equation}%
where $\mathcal{L}_{m}$ is the Lagrangian of dust-like ($p_{m}=0$) matter
and $k^{2}=8\pi G$. Varying the action with respect to the metric\footnote{%
We use the metric i.e. the Hilbert variational approach.} we arrive at
\begin{equation}
(1+f^{\prime })G_{\nu }^{\mu }\,-\,g^{\mu \alpha }f_{_{R},\,\alpha \,;\,\nu
}+\left[ \frac{2\Box f^{\prime }-(f-Rf^{\prime })}{2}\right] \delta _{\;\nu
}^{\mu }=k^{2}\,T_{\nu }^{\mu }  \label{EE}
\end{equation}%
where the prime denotes derivative with respect to $R$, $G_{\nu }^{\mu }$ is
the Einstein tensor and $T_{\nu }^{\mu }$ is the ordinary energy-momentum
tensor of matter. Based on the matter era we treat the expanding universe as
a perfect fluid which includes only cold dark matter with comoving observers
$U^{\mu }=\delta _{0}^{\mu }$. Thus the energy momentum tensor becomes $%
T_{\mu \nu }=\rho _{m}U_{\mu }U_{\nu }$, where $\rho _{m}$ is the energy
density of the cosmic fluid.

Now, in the context of a flat FRW model the metric is
\begin{equation}
ds^{2}=-dt^{2}+a^{2}(t)(dx^{2}+dy^{2}+dz^{2}).  \label{SF.1}
\end{equation}%
The components of the Einstein tensor are computed to be:
\begin{equation}
G_{0}^{0}=-3H^{2},\;G_{b}^{a}=-\delta _{b}^{a}\left( 2\dot{H}+3H^{2}\right)
\;.  \label{EIN.1}
\end{equation}%
Inserting (\ref{EIN.1}) into the modified Einstein's field equations (\ref%
{EE}), for comoving observers, we derive the modified Friedman's equation
\begin{equation}
3f^{^{\prime }}H^{2}=k^{2}\rho _{m}+\frac{f^{^{\prime }}R-f}{2}%
-3Hf^{^{\prime \prime }}\dot{R}  \label{motion1}
\end{equation}

\begin{equation}
2f^{^{\prime }}\dot{H}+3f^{^{\prime }}H^{2}=-2Hf^{^{\prime \prime }}\dot{R}%
-\left( f^{^{\prime \prime \prime }}\dot{R}^{2}+f^{^{\prime \prime }}\ddot{R}%
\right) -\frac{f-Rf^{^{\prime }}}{2}.  \label{motion2}
\end{equation}%
The contraction of the Ricci tensor provides the Ricci scalar
\begin{equation}
R=g^{\mu \nu }R_{\mu \nu }=6\left( \frac{\ddot{a}}{a}+\frac{\dot{a}^{2}}{%
a^{2}}\right) =6(2H^{2}+\dot{H})\;.  \label{SF.3b}
\end{equation}%
The Bianchi identity $\bigtriangledown ^{\mu }\,{T}_{\mu \nu }=0$ leads to
the matter conservation law:%
\begin{equation}
\dot{\rho}_{m}+3H\rho _{m}=0\,  \label{frie3}
\end{equation}%
whose solution is
\begin{equation}
\rho _{m}=\rho _{m0}a^{-3}.
\end{equation}%
Note that the over-dot denotes derivative with respect to the cosmic time $t$
and $H\equiv \dot{a}/a$ is the Hubble parameter.

If we consider $f(R)=R$ then the field equations (\ref{EE}) boil down to the
Einstein's equations a solution of which is the Einstein de Sitter model. On
the other hand, the concordance $\Lambda $ cosmology is fully recovered for $%
f(R)=R-2\Lambda $.

From the current analysis it becomes clear that unlike the standard Friedman
equations in Einstein's GR the modified equations of motion (\ref{motion1})
and (\ref{motion2}) are complicated and thus it is difficult to solve them
analytically.

We would like to stress here that within the context of the metric formalism
the above $f(R)$ cosmological models must obey simultaneously some strong
conditions \cite{AmenB}. These are: (i) $f^{^{\prime }}>0$ for $R\geq
R_{0}>0 $, where $R_{0}$ is the Ricci scalar at the present time. If the
final attractor is a de Sitter point we need to have $f^{^{\prime }}>0$ for $%
R\geq R_{1}>0$, where $R_{1}$ is the Ricci scalar at the de Sitter point,
(ii) $f^{^{\prime \prime }}>0$ for $R\geq R_{0}>0$, (iii) $f(R)\approx
R-2\Lambda $ for $R\gg R_{0}$ and finally (iv) $0<\frac{Rf^{^{\prime \prime
}}}{f^{^{\prime }}}(r)<1$ at $r=-\frac{Rf^{^{\prime }}}{f}=-2$

\section{Modified gravity versus symmetries}

\label{MGvsSy}

In the last decade a large number of experiments have been proposed in order
to constrain dark energy and study its evolution. Naturally, in order to
establish the evolution of the dark energy (\textquotedblright
geometrical\textquotedblright\ in the current work) equation of state
parameter a realistic form of $H(a)$ is required while the included free
parameters must be constrained through a combination of independent DE
probes (for example SNIa, BAOs, CMB etc). However, a weak point here is the
fact that the majority of the $f(R)$ models appeared in the literature are
plagued with no clear physical basis and/or many free parameters. Due to the
large number of free parameters many such models could fit the data. The
proposed additional criterion of Noether point symmetry requirement is a
physically meaning-full geometric ansatz.

According to the theory of general relativity, the space-time symmetries
(Killing and homothetic vectors) via the Einstein's field equations, are
also symmetries of the energy momentum tensor. Due to the fact that the $%
f(R) $ models provide a natural generalization of GR one would expect that
the theories of modified gravity must inherit the symmetries of the
space-time as the usual gravity (GR) does.

Furthermore, besides the geometric symmetries we have to consider the
dynamical symmetries, which are the symmetries of the field equations (Lie
symmetries). If the field equations are derived from a Lagrangian then there
is a special class of Lie symmetries, the Noether symmetries, which lead to
conserved currents or, equivalently, to first integrals of the equations of
motion. The Noether integrals are used to reduce the order of the field
equations or even to solve them. Therefore a sound requirement, which is
possible to be made in Lagrangian theories, is that they admit extra Noether
symmetries. This assumption is model independent, because it is imposed
after the field equations have been derived, therefore it does not lead to
conflict with the geometric symmetries while, at the same time, serves the
original purpose of a selection rule. Of course, it is possible that a
different method could be assumed and select another subset of viable
models. However, symmetry has always played a dominant role in Physics and
this gives an aesthetic and a physical priority to our proposal.

In the Lagrangian context, the main field equations (\ref{motion1}) and (\ref%
{motion2}), described in section \ref{CMG}, can be produced by the following
Lagrangian:
\begin{equation}
L\left( a,\dot{a},R,\dot{R}\right) =6af^{^{\prime }}~\dot{a}%
^{2}+6a^{2}f^{^{\prime \prime }}~\dot{a}\dot{R}+a^{3}\left( f^{^{\prime
}}R-f\right) \qquad  \label{SF.50}
\end{equation}%
in the space of the variables $\{a,R\}$. Using eq.(\ref{SF.50}) we obtain
the Hamiltonian of the current dynamical system
\begin{equation}
E=6af^{^{\prime }}~\dot{a}^{2}+6a^{2}f^{^{\prime \prime }}~\dot{a}\dot{R}%
-a^{3}\left( f^{^{\prime }}R-f\right)  \label{SF.60e}
\end{equation}%
or
\begin{equation}
E=6a^{3}\left[ f^{^{\prime }}H^{2}-\frac{1}{6f^{^{\prime }}}\left( \left(
f^{^{\prime }}R-f\right) -6\dot{R}Hf^{^{\prime \prime }}\right) \right] \;.
\label{SF.61e}
\end{equation}%
Combining the first equation of motion (\ref{motion1}) with eq.(\ref{SF.61e}%
) we find
\begin{equation}
\rho _{m}=\frac{E}{2k^{2}}\;a^{-3}\;.  \label{Smm}
\end{equation}%
The latter equation together with $\rho _{m}=\rho _{m0}a^{-3}$ implies that
\begin{equation}
\rho _{m0}=\frac{E}{2k^{2}}\Rightarrow \Omega _{m}\rho _{cr,0}=\frac{E}{%
2k^{2}}\Rightarrow E=6\Omega _{m}H_{0}^{2}
\end{equation}%
where $\Omega _{m}=\rho _{m0}/\rho _{cr,0}$, $\rho _{cr,0}=3H_{0}^{2}/k^{2}$
is the critical density at the present time and $H_{0}$ is the Hubble
constant.

We note that the current Lagrangian eq.(\ref{SF.50}) is time independent
implying that the dynamical system is autonomous hence the Hamiltonian $E$
is conserved ($\frac{dE}{dt}=0$).

\section{Noether symmetries}

Before we proceed we review briefly the basic definitions concerning Lie and
Noether point symmetries of systems of second order ordinary differential
equations (ODEs)%
\begin{equation}
\ddot{x}^{i}=\omega ^{i}\left( t,x^{j},\dot{x}^{j}\right) .  \label{Lie.0}
\end{equation}%
The one point parameter transformation%
\begin{eqnarray}
\bar{t} &=&t+\varepsilon \xi \left( t,x^{i}\right) \\
\bar{x}^{i} &=&x^{i}+\varepsilon \eta ^{i}\left( t,x^{i}\right)
\end{eqnarray}%
with generator $X=\xi \left( t,x^{j}\right) \partial _{t}+\eta ^{i}\left(
t,x\right) \partial _{i}$\ is a Lie point symmetry of the system of ODEs (%
\ref{Lie.0})\ if the following condition is satisfied \cite{Olver,StephaniB}
\begin{equation}
X^{\left[ 2\right] }\left( \ddot{x}^{i}-\omega \left( t,x^{j},\dot{x}%
^{j}\right) \right) =0  \label{Lie.1}
\end{equation}%
where $X^{\left[ 2\right] }$\ is the second prolongation of $X$ defined by
the formula%
\begin{equation}
X^{\left[ 2\right] }=\xi \partial _{t}+\eta ^{i}\partial _{i}+\left( \dot{%
\eta}^{i}-\dot{x}^{i}\dot{\xi}\right) \partial _{\dot{x}^{i}}+\left( \ddot{%
\eta}^{i}-\dot{x}^{i}\ddot{\xi}-2\ddot{x}^{i}\dot{\xi}\right) \partial _{%
\ddot{x}^{i}}.  \label{Lie.2}
\end{equation}%
Condition (\ref{Lie.1}) is equivalent to the relation
\begin{equation}
\left[ X^{\left[ 1\right] },A\right] =\lambda \left( x^{a}\right) A
\label{Lie.3a}
\end{equation}%
where $X^{\left[ 1\right] }~$is the first prolongation of $X$ and $A$ is the
Hamiltonian vector field%
\begin{equation}
A=\partial _{t}+\dot{x}\partial _{x}+\omega ^{i}\left( t,x^{j},\dot{x}%
^{j}\right) \partial _{\dot{x}^{i}}.  \label{Lie.4}
\end{equation}

If the system of ODEs results from a first order Lagrangian $L=L\left(
t,x^{j},\dot{x}^{j}\right) ,$ then a Lie symmetry $X$ of the system (\ref%
{Lie.0}) is a Noether symmetry of the Lagrangian if the additional condition
is satisfied
\begin{equation}
X^{\left[ 1\right] }L+L\frac{d\xi }{dt}=\frac{dg}{dt}  \label{Lie.5}
\end{equation}%
where $g=g\left( t,x^{j}\right) $\ is the gauge function. To every Noether
symmetry there corresponds a first integral (a Noether integral) of the
system of equations (\ref{Lie.0}) which is given by the formula:%
\begin{equation}
I=\xi E_{H}-\frac{\partial L}{\partial \dot{x}^{i}}\eta ^{i}+g  \label{Lie.6}
\end{equation}%
where $E_{H}$ is the Hamiltonian of the Lagrangian
\begin{equation}
E_{H}=\dot{x}^{i}\frac{\partial L}{\partial x^{i}}-L  \label{Lie.7}
\end{equation}

The vector field $X$ in the augmented space $\{t,a,R\}$ is
\begin{equation}
X=\xi \left( t,a,R\right) \partial _{t}+\eta ^{\left( 1\right) }\left(
t,a,R\right) \partial _{a}+\eta ^{\left( 2\right) }\left( t,a,R\right)
\partial _{R}  \label{Lie.3}
\end{equation}%
and the first prolongation%
\begin{equation}
X^{\left[ 1\right] }=\xi \partial _{t}+\eta ^{\left( 1\right) }\partial
_{a}+\eta ^{\left( 2\right) }\partial _{R}+\left( \eta ^{\left( 1\right) }-%
\dot{a}\dot{\xi}\right) \partial _{\dot{a}}+\left( \dot{\eta}^{\left(
2\right) }-\dot{R}\dot{\xi}\right) \partial _{R}.
\end{equation}

Having given the basic formula for the Noether symmetries we look for
analytic solutions of the dynamical system with Lagrangian (\ref{SF.50})
with the use of Noether Integrals.

\section{Noether symmetries of $f\left( R\right) $ gravity}

\label{LieN}

The Noether condition (\ref{Lie.5}) for the Lagrangian (\ref{SF.50}) is
equivalent with the following system of eight equations
\begin{eqnarray}
\xi _{,a} &=&0~  \label{NC.01} \\
~\xi _{,R} &=&0  \label{NC.02}
\end{eqnarray}%
\begin{equation}
a^{2}f^{\prime \prime }\eta _{,R}^{\left( 1\right) }=0  \label{NC.03}
\end{equation}%
\begin{equation}
f^{\prime }\eta ^{\left( 1\right) }+af^{\prime \prime }\eta ^{\left(
2\right) }+2af^{\prime }\eta _{,a}^{\left( 1\right) }+a^{2}f^{\prime \prime
}\eta _{,a}^{\left( 2\right) }-\frac{1}{2}af^{\prime }\xi _{,t}=0
\label{NC.04}
\end{equation}%
\begin{equation}
2af^{\prime \prime }\eta ^{\left( 1\right) }+a^{2}f^{\prime \prime \prime
}\eta ^{\left( 2\right) }+a^{2}f^{\prime \prime }\eta _{,a}^{\left( 1\right)
}+2af^{\prime }\eta _{,R}^{\left( 1\right) }+a^{2}f^{\prime \prime }\eta
_{,R}^{\left( 2\right) }-\frac{1}{2}a^{2}f^{\prime \prime }\xi _{,t}=0
\label{NC.05}
\end{equation}%
\begin{equation}
-3a^{2}Rf^{\prime }\eta ^{\left( 1\right) }+3a^{2}f\eta ^{\left( 1\right)
}-a^{3}Rf^{\prime \prime }\eta ^{\left( 2\right) }+a^{3}\left( f-f^{^{\prime
}}R\right) \xi _{,t}+g_{,t}=0  \label{NC.06}
\end{equation}

\begin{equation}
12af^{\prime }\eta _{,t}^{\left( 1\right) }+6a^{2}f^{\prime \prime }\eta
_{,t}^{\left( 2\right) }+a^{3}\left( f^{^{\prime }}R-f\right) \xi
_{,a}-g_{,a}=0  \label{NC.07}
\end{equation}%
\begin{equation}
6a^{2}f^{\prime \prime }\eta _{,t}^{\left( 1\right) }+a^{3}\left(
f^{^{\prime }}R-f\right) \xi _{,R}-g_{,R}=0  \label{NC.08}
\end{equation}%
The solution of the system (\ref{NC.01})-(\ref{NC.08}) will determine the
Noether symmetries.

Since the Lagrangian (\ref{SF.50}) is in the form $L=T\left( a,\dot{a},R,%
\dot{R}\right) -V\left( a,R\right) $, the results of \cite{Tsam10} can be
used \footnote{%
Where $T$ is the "kinetic" term and $V$ is the "potential}. The kinematic
term defines a two dimensional metric in the space of $\{a,R\}$ with line
element%
\begin{equation}
d\hat{s}^{2}=12af^{\prime }da^{2}+12a^{2}f^{\prime \prime }da~dR
\label{FR.03}
\end{equation}%
while the \textquotedblright potential\textquotedblright\ is
\begin{equation}
V(a,R)=-a^{3}(f^{^{\prime }}R-f)\;.  \label{pot}
\end{equation}

The Ricci scalar of the two dimensional metric (\ref{FR.03}) is computed to
be $\hat{R}=0,$ therefore the space is a flat space\footnote{%
All two dimensional Riemannian spaces are Einstein spaces implying that if $%
\hat{R}=const$ the space is maximally symmetric \cite{Barnes1993} and if $%
\hat{R}=0,$ the space admit gradient Homothetic vector, i.e. is flat.} with
a maximum homothetic algebra. \ The homothetic algebra of the metric (\ref%
{FR.03}) consists of the vectors
\begin{align*}
\mathbf{K}^{1}& =a\partial _{a}-3\frac{f^{\prime }}{f^{\prime \prime }}%
\partial _{R}~,~\mathbf{K}^{2}=\frac{1}{a}\partial _{a}-\frac{1}{a^{2}}\frac{%
f^{\prime }}{f^{\prime \prime }}\partial _{R}~ \\
\mathbf{K}^{3}& =\frac{1}{a}\frac{1}{f^{\prime \prime }}\partial _{R}~,~%
\mathbf{H}=\frac{a}{2}~\partial _{a}+\frac{1}{2}\frac{f^{\prime }}{f^{\prime
\prime }}\partial _{R}
\end{align*}%
where $\mathbf{K}$ are Killing vectors ($\mathbf{K}^{2,3}$ are gradients)
and $\mathbf{H}$ is a gradient Homothetic vector.

Therefore applying theorem 2 of \cite{Tsam10} we have the following cases:

\textbf{Case 1:} If $f\left( R\right) $ is arbitrary the dynamical system
admits as Noether symmetry~the~$X^{1}=\partial _{t}~$with Noether integral
the Hamiltonian~$E$.

\textbf{Case 2:} If $f\left( R\right) =R^{\frac{3}{2}}~$the dynamical system
admits the extra Noether symmetries%
\begin{equation}
X^{2}=\mathbf{K}^{2},~X^{3}=t\mathbf{K}^{2}  \label{NS.01}
\end{equation}%
\begin{equation}
X^{4}=2t\partial _{t}+\mathbf{H+}\frac{5}{6}\mathbf{K}^{1}.  \label{NS.02}
\end{equation}%
with corresponding Noether Integrals
\begin{equation}
I_{2}=\frac{d}{dt}\left( a\sqrt{R}\right)  \label{NI.02}
\end{equation}%
\begin{equation}
I_{3}=t\frac{d}{dt}\left( a\sqrt{R}\right) -a\sqrt{R}  \label{NI.03}
\end{equation}%
\begin{equation}
I_{4}=2tE-6a^{2}\dot{a}\sqrt{R}-6\frac{a^{3}}{\sqrt{R}}\dot{R}.
\label{NI.04}
\end{equation}%
the non vanishing commutators of the Noether algebra are%
\begin{equation*}
\left[ X^{1},X^{3}\right] =X^{2}~~~~\left[ X^{1},X^{4}\right] =2X^{1}
\end{equation*}%
\begin{equation*}
\left[ X^{2},X^{4}\right] =\frac{8}{3}X^{2}~~~~\left[ X^{3},X^{4}\right] =%
\frac{2}{3}X^{3}
\end{equation*}

\textbf{Case 3:} If $f\left( R\right) =R^{\frac{7}{8}}$ the dynamical system
admits the extra Noether symmetries%
\begin{equation}
X^{5}=2t\partial _{t}+\mathbf{H}~,~X^{6}=t^{2}\partial _{t}+t\mathbf{H}
\label{NS.03}
\end{equation}%
with corresponding Noether Integrals
\begin{equation}
I_{5}=2tE-\frac{21}{8}\frac{d}{dt}\left( a^{3}R^{-\frac{1}{8}}\right)
\label{NI.05}
\end{equation}%
\begin{equation}
I_{6}=t^{2}E-\frac{21}{8}t\frac{d}{dt}\left( a^{3}R^{-\frac{1}{8}}\right) +%
\frac{21}{8}a^{3}R^{-\frac{1}{8}}.  \label{NI.06}
\end{equation}%
\qquad and the non vanishing commutators of the Noether algebra are%
\begin{equation*}
\left[ X^{1},X^{5}\right] =2X^{1}~~~~\left[ X^{1},X^{6}\right] =X^{5}~~~~%
\left[ X^{5},X^{6}\right] =2X^{6}
\end{equation*}%
From the time dependent integrals (\ref{NI.05}),(\ref{NI.06}) and the
Hamiltonian we construct the Ermakov-Lewis invariant \cite{MoyoLeach,Ermakov}
\begin{equation}
\Sigma =4I_{6}E-I_{5}^{2}  \label{NI.06b}
\end{equation}

\textbf{Case 4}: If $f\left( R\right) =\left( R-2\Lambda \right) ^{\frac{3}{2%
}}$ the dynamical system admits the extra Noether symmetries
\begin{equation}
\bar{X}^{2}=e^{\sqrt{m}t}\mathbf{K}^{2}~,~\bar{X}^{3}=e^{-\sqrt{m}t}\mathbf{K%
}^{2}  \label{NS.b2}
\end{equation}%
with corresponding Noether Integrals
\begin{equation}
\bar{I}_{2}=e^{\sqrt{m}t}\left( \frac{d}{dt}\left( a\sqrt{R-2L}\right) -9%
\sqrt{m}a\sqrt{R-2\Lambda }\right)   \label{NI.b2}
\end{equation}%
\begin{equation}
\bar{I}_{3}=e^{-\sqrt{m}t}\left( \frac{d}{dt}\left( a\sqrt{R-2L}\right) +9%
\sqrt{m}a\sqrt{R-2\Lambda }\right)   \label{NI.b3}
\end{equation}%
where $m=\frac{2}{3}\Lambda .$ The non vanishing commutators of the Noether
algebra are%
\begin{equation*}
\left[ X^{1},\bar{X}^{2}\right] =\sqrt{m}\bar{X}^{2}~~~~\left[ \bar{X}%
^{3},X^{1}\right] =\sqrt{m}\bar{X}^{3}~
\end{equation*}%
From the time dependent integrals (\ref{NI.b2}),(\ref{NI.b3}) we construct
the time independent integral $\bar{I}_{23}=\bar{I}_{2}\bar{I}_{3}.$

\textbf{Case 5}: If $f\left( R\right) =\left( R-2\Lambda \right) ^{\frac{7}{8%
}}$ the dynamical system admits the extra Noether symmetries%
\begin{eqnarray}
\bar{X}^{5} &=&\frac{1}{\sqrt{m}}e^{2\sqrt{m}t}\partial _{t}+e^{2\sqrt{m}t}~%
\mathbf{H}  \label{NS.b5} \\
\bar{X}^{6} &=&-\frac{1}{\sqrt{m}}e^{-2\sqrt{m}t}\partial _{t}+e^{-2\sqrt{m}%
t}~\mathbf{H}  \label{NS.b6}
\end{eqnarray}%
with corresponding Noether Integrals
\begin{equation}
\bar{I}_{5}=e^{2\sqrt{m}t}\left[ \frac{1}{\sqrt{m}}E-\frac{21}{8}\frac{d}{dt}%
\left( a^{3}\left( R-2\Lambda \right) ^{-\frac{1}{8}}\right) +\frac{21}{4}%
\sqrt{m}a^{3}\left( R-2\Lambda \right) ^{-\frac{1}{8}}\right]   \label{NI.b5}
\end{equation}%
\begin{equation}
\bar{I}_{6}=e^{-2\sqrt{m}t}\left[ \frac{1}{\sqrt{m}}E+\frac{21}{8}\frac{d}{dt%
}\left( a^{3}\left( R-2\Lambda \right) ^{-\frac{1}{8}}\right) +\frac{21}{4}%
\sqrt{m}a^{3}\left( R-2\Lambda \right) ^{-\frac{1}{8}}\right]   \label{NI.b6}
\end{equation}%
and the non vanishing commutators of the Noether algebra are%
\begin{equation*}
\left[ X^{1},\bar{X}^{5}\right] =2\sqrt{m}\bar{X}^{5}~~~~\left[ \bar{X}%
^{6},X^{1}\right] =2\sqrt{m}\bar{X}^{6}
\end{equation*}%
\begin{equation*}
\left[ \bar{X}^{5},\bar{X}^{6}\right] =\frac{4}{\sqrt{m}}X^{1}
\end{equation*}%
From the time dependent integrals (\ref{NI.05}),(\ref{NI.06}) and the
Hamiltonian we construct the Ermakov-Lewis invariant \cite{Ermakov}
\begin{equation}
\phi =E^{2}-\bar{I}_{5}\bar{I}_{6}  \label{NI.b6b}
\end{equation}

\textbf{Case 6:~}If $f\left( R\right) =R^{n}$ (with $n\neq 0,1,\frac{3}{2},%
\frac{7}{8}$) the dynamical system admits the extra Noether symmetry%
\begin{equation}
X^{7}=2t\partial _{t}+\mathbf{H+}\left( \frac{4n}{3}-\frac{7}{6}\right)
\mathbf{K}^{1}  \label{NS.07}
\end{equation}%
with corresponding Noether Integral
\begin{equation}
I_{7}=2tE-8na^{2}R^{n-1}\dot{a}\left( 2-n\right) -4na^{3}R^{n-2}\dot{R}%
\left( 2n-1\right) \left( n-1\right) .  \label{NI.07}
\end{equation}%
and the commutator of the Noether algebra is $\left[ X^{1},X^{7}\right]
=2X^{1}.$

We note that the Noether subalgebra of case 2, $\left\{
X^{1},X^{2},X^{3}\right\} $ and the algebra of case 4 $\left\{ X^{1},\bar{X}%
^{2},\bar{X}^{3}\right\} $ is the same Lie algebra but not in the same
representation. The same observation applies to the subalgebra of case 3 $%
\left\{ X^{1},X^{5},X^{6}\right\} $ and the algebra of case 5 $\left\{ X^{1},%
\bar{X}^{5},\bar{X}^{6}\right\} $. This connection between the Lie groups is
useful because it reveals common features in the dynamic systems, as is the
common transformation to the normal coordinates of the systems.

For the cosmological viability of the models see \cite%
{AmeTsuj,Amenp,Paliathanasis}

\section{Analytic Solutions}

\label{Analsol}

Using the Noether symmetries and the associated Noether integrals we solve
analytically the differential eqs.(\ref{motion1}), (\ref{motion2}) and (\ref%
{SF.3b}) for the cases where the dynamical system is Liouville integrable,
that is for cases 2-5. Case 6$~$(i.e. $f\left( R\right) =R^{n}$) is not
Liouville integrable via Noether point symmetries, since the Noether
integral (\ref{NI.07}) is time dependent\footnote{%
In the appendix \ref{AppendixA} we present special solutions for the $%
f\left( R\right) =R^{n}~$model, using the zero order invariants.}.

\subsection{Power law model $R^{\protect\mu }$ with $\protect\mu =\frac{3}{2}
$}

\label{subs1}

In this case the Lagrangian eq.(\ref{SF.50}) of the $f(R)=R^{\frac{3}{2}}$
model is written as
\begin{equation}
L=9a\sqrt{R}\dot{a}^{2}+\frac{9a^{2}}{2\sqrt{R}}\dot{a}\dot{R}+\frac{a^{3}}{2%
}R^{\frac{3}{2}}  \label{FR.32}
\end{equation}%
Changing the variables from $(a,R)$ to $(z,w)$ via the relations:
\begin{equation}
a=\left( \frac{9}{2}\right) ^{-\frac{1}{3}}\sqrt{z}\;\;\;\;R=\frac{w^{2}}{z}
\end{equation}%
the Lagrangian (\ref{FR.32}) and the Hamiltonian (\ref{SF.60e}) become
\begin{equation}
L=\dot{z}\dot{w}+V_{0}w^{3}
\end{equation}%
\begin{equation}
E=\dot{z}\dot{w}-V_{0}w^{3}
\end{equation}%
where $V_{0}=\frac{1}{9}.~$The equations of motion in the new coordinate
system are
\begin{eqnarray}
\ddot{w} &=&0 \\
\ddot{z}-3V_{0}w^{2} &=&0
\end{eqnarray}%
The Noether integrals (\ref{NI.02}),(\ref{NI.03}) in the coordinate system $%
\left\{ z,y\right\} $ are
\begin{equation}
I_{1}^{^{\prime }}=\dot{w}~\ ,~~I_{2}^{^{\prime }}=t\dot{w}-w
\end{equation}%
The general solution of the system is:
\begin{equation}
y\left( t\right) =I_{1}^{\prime }t-I_{2}^{^{\prime }}
\end{equation}%
\begin{equation}
z\left( t\right) =\frac{1}{36\left( I_{1}^{^{\prime }}\right) ^{2}}\left(
I_{1}^{^{\prime }}t-I_{2}^{^{\prime }}\right) ^{4}+z_{1}t+z_{0}
\end{equation}%
The Hamiltonian constrain gives~$E=z_{1}I_{1}^{^{\prime }}$ where $z_{0,1}~$%
are constants and the singularity condition results in the constrain%
\begin{equation}
\frac{1}{36\left( I_{1}^{^{\prime }}\right) ^{2}}\left( I_{2}^{^{\prime
}}\right) ^{4}+z_{0}=0.
\end{equation}

\subsection{Power law model $R^{\protect\mu }$ with $\protect\mu =\frac{7}{8}
$}

\label{subsb1}

In this case the Lagrangian eq.(\ref{SF.50}) is written as
\begin{equation}
L=\frac{21a}{4R^{\frac{1}{8}}}\dot{a}^{2}-\frac{21}{16}\frac{a^{2}}{R^{\frac{%
9}{8}}}\dot{a}\dot{R}-\frac{1}{8}a^{3}R^{\frac{7}{8}}.  \label{FR.78}
\end{equation}%
Changing now the variables from $(a,R)$ to $(\rho ,\sigma )$ via the
relations:
\begin{equation}
a=\left( \frac{21}{4}\right) ^{-\frac{1}{3}}\sqrt{\rho e^{\sigma }}\;\;\;\;R=%
\frac{e^{12\sigma }}{\rho ^{4}}.
\end{equation}%
The Lagrangian (\ref{FR.16}) and the Hamiltonian (\ref{SF.60e}) become%
\begin{equation}
L=\frac{1}{2}\dot{\rho}^{2}-\frac{1}{2}\rho ^{2}\dot{\sigma}^{2}+V_{0}\frac{%
e^{12\sigma }}{\rho ^{2}}  \label{FR.78a}
\end{equation}%
\begin{equation}
E=\frac{1}{2}\dot{\rho}^{2}-\frac{1}{2}\rho ^{2}\dot{\sigma}^{2}-V_{0}\frac{%
e^{12\sigma }}{\rho ^{2}}.  \label{FR.78b}
\end{equation}%
where$~V_{0}=-\frac{1}{42}.$ \ The Euler-Lagrange equations provide the
following equations of motion:%
\begin{align}
\ddot{\rho}+\rho \dot{\sigma}^{2}+2V_{0}\frac{e^{12\sigma }}{\rho ^{3}}& =0
\label{FR.78c} \\
\ddot{\sigma}+\frac{2}{\rho }\dot{\sigma}\dot{\rho}+12V_{0}\frac{e^{12\sigma
}}{\rho ^{2}}& =0.  \label{FR.78d}
\end{align}%
The Noether integrals (\ref{NI.05}), (\ref{NI.06}) and the Ermakov-Lewis
invariant \ref{NI.06b} in the coordinate system $\left\{ u,v\right\} $ are
\begin{align}
I_{5}^{^{\prime }}& =2tE-\rho \dot{\rho} \\
I_{6}^{^{\prime }}& =t^{2}E-t\rho \dot{\rho}+\frac{1}{2}\rho ^{2}.
\end{align}%
\begin{equation}
\Sigma =\rho ^{4}\dot{\sigma}^{2}+4V_{0}e^{12\sigma }.
\end{equation}%
Using the Ermakov-Lewis Invariant, the Hamiltonian (\ref{FR.78a}) and
equation (\ref{FR.78c}) are written:%
\begin{align}
\frac{1}{2}\dot{\rho}^{2}-\frac{1}{2}\frac{\Sigma }{\rho ^{2}}& =E \\
\ddot{\rho}+\frac{\Sigma }{\rho ^{3}}& =0.
\end{align}%
And the analytical solution of the system is
\begin{equation}
\rho \left( t\right) =\left( \rho _{2}t^{2}+\rho _{1}t+\frac{\left( \left(
\rho _{1}\right) ^{2}-4\Sigma \right) }{4\rho _{2}}\right) ^{\frac{1}{2}}
\end{equation}%
\begin{equation}
\exp \left( \sigma \left( t\right) \right) =\left\{ \frac{21}{2}\Sigma \left[
\left( \tanh \left[ \sigma _{0}\rho _{2}\sqrt{\Sigma }-6\arctan h\left(
\frac{2\rho _{2}t+\rho _{1}}{2\sqrt{\Sigma }}\right) \right] \right) ^{2}-1%
\right] \right\} ^{\frac{1}{12}}
\end{equation}%
where $B\left( t\right) =\left( \frac{1}{2}\frac{2\rho _{2}t+\rho _{1}}{%
\sqrt{\Sigma }}\right) $ and $\rho _{1,2}~$,$~\sigma _{0}$ are constants
with Hamiltonian constrain $E=\frac{1}{2}\rho _{2}~$. The singularity
constrain gives $\left( \rho _{1}\right) ^{2}=4\Sigma $

In the case $\Sigma =0$ the analytical solution is%
\begin{equation}
\rho \left( t\right) =\left( \rho _{2}t^{2}+\rho _{1}t+\frac{1}{2}\frac{%
\left( \rho _{1}\right) ^{2}}{\rho _{2}}\right) ^{\frac{1}{2}}
\end{equation}%
\begin{equation}
\exp \sigma \left( t\right) =\left[ \frac{1}{24\sqrt{V_{0}}}\frac{\left(
2\rho _{2}t+\rho _{1}\right) }{\left( 4\sigma _{0}\rho _{2}^{2}t+2\sigma
_{0}\rho _{2}\rho _{1}-1\right) }\right] ^{\frac{1}{6}}
\end{equation}%
The singularity constrain gives $\rho _{1}=0$, then the solution is%
\begin{equation}
a\left( t\right) =\frac{a_{0}t^{\frac{7}{6}}}{\left( a_{2}t-1\right) ^{\frac{%
1}{6}}}
\end{equation}

In contrast with the claim of \cite{Nayem12} this model is analytically
solvable and there exists models which admit Noether integrals with time
dependent gauge functions.

\subsection{$\Lambda _{bc}$CDM model with $(b,c)=(1,\frac{3}{2})$}

\label{subs}

Inserting $f(R)=(R-2\Lambda )^{3/2}$ into eq.(\ref{SF.50}) we obtain
\begin{equation}
L=9a\sqrt{R-2\Lambda }\dot{a}^{2}+\frac{9a^{2}}{2\sqrt{R-2\Lambda }}\dot{a}%
\dot{R}+\frac{a^{3}}{2}\left( R+4\Lambda \right) \sqrt{R-2\Lambda }
\label{FR.13a}
\end{equation}%
Changing now the variables from $(a,R)$ to $(x,y)$ via the relations:
\begin{equation}
a=\left( \frac{9}{2}\right) ^{-\frac{1}{3}}\sqrt{x}\;\;\;\;R=2\Lambda +\frac{%
y^{2}}{x}
\end{equation}%
the Lagrangian (\ref{FR.13a}) and the Hamiltonian (\ref{SF.60e}) become
\begin{equation}
L=\dot{x}\dot{y}+V_{0}\left( y^{3}+\bar{m}xy\right)
\end{equation}%
\begin{equation}
E=\dot{x}\dot{y}-V_{0}\left( y^{3}+{\bar{m}}xy\right)  \label{hamm}
\end{equation}%
where $V_{0}=\frac{1}{9}$ and $\bar{m}=6\Lambda $. \newline
The equations of motion, using the Euler-Lagrange equations, in the new
coordinate system are
\begin{equation}
\ddot{x}-3V_{0}y^{2}-{\bar{m}V}_{0}x=0  \label{FR.13b}
\end{equation}%
\begin{equation}
\ddot{y}-{\bar{m}}V_{0}y=0.  \label{FR.14}
\end{equation}%
The Noether integrals (\ref{NI.b2}),(\ref{NI.b3}) in the coordinate system $%
\left\{ x,y\right\} $ are
\begin{align}
\bar{I}_{1}^{\prime }& =e^{\omega t}\dot{y}-\omega e^{\omega t}y
\label{II.14} \\
\bar{I}_{2}^{\prime }& =e^{-\omega t}\dot{y}+\omega e^{-\omega t}y.
\label{II1.14}
\end{align}%
where $\omega =\sqrt{2\Lambda /3}$. From these we construct the time
independent first integral%
\begin{equation}
\Phi =I_{1}I_{2}=\dot{y}^{2}-\omega ^{2}y^{2}.  \label{II2.14}
\end{equation}%
The constants of integration are further constrained by the condition that
at the singularity ($t=0$), the scale factor has to be exactly zero, that
is, $x(0)=0$.\newline
The general solution of the system (\ref{FR.13b})-(\ref{FR.14}) is:
\begin{equation}
y\left( t\right) =\frac{I_{2}}{2\omega }e^{\omega t}-\frac{I_{1}}{2\omega }%
e^{-\omega t}
\end{equation}%
\begin{equation}
x\left( t\right) =x_{1G}e^{\omega t}+x_{2G}e^{-\omega t}+\frac{1}{4\bar{m}%
\omega ^{2}}\left( I_{2}e^{\omega t}+I_{1}e^{-\omega t}\right) ^{2}+\frac{%
\Phi }{\bar{m}\omega ^{2}}.
\end{equation}%
The Hamiltonian constrain gives~$E=\omega \left(
x_{1G}I_{1}-x_{2G}I_{2}\right) $ where $x_{1G,2G}~$are constants and the
singularity condition results in the constrain%
\begin{equation}
x_{1G}+x_{2G}+\frac{1}{4\bar{m}\omega ^{2}}\left( I_{1}+I_{2}\right) ^{2}+%
\frac{\Phi }{\bar{m}\omega ^{2}}=0.
\end{equation}%
At late enough times the solution becomes $a^{2}(t)\propto e^{2\omega t}$

\subsection{$\Lambda _{bc}$CDM model with $(b,c)=(1,\frac{7}{8})$}

\label{subsb2}

In this case the Lagrangian eq.(\ref{SF.50}) of the $f(R)=(R-2\Lambda
)^{7/8} $ model is written as
\begin{equation}
L=\frac{21a}{4\left( R-2\Lambda \right) ^{\frac{1}{8}}}\dot{a}^{2}-\frac{21}{%
16}\frac{a^{2}}{\left( R-2\Lambda \right) ^{\frac{9}{8}}}\dot{a}\dot{R}-%
\frac{1}{8}a^{3}\frac{\left( R-16\Lambda \right) }{\left( R-2\Lambda \right)
^{\frac{1}{8}}}.  \label{FR.16}
\end{equation}%
Changing now the variables from $(a,R)$ to $(u,v)$ via the relations:
\begin{equation}
a=\left( \frac{21}{4}\right) ^{-\frac{1}{3}}\sqrt{ue^{v}}\;\;\;\;R=2\Lambda +%
\frac{e^{12v}}{u^{4}}.  \label{FR.16a}
\end{equation}%
The Lagrangian (\ref{FR.16}) and the Hamiltonian (\ref{SF.60e}) become%
\begin{equation}
L=\frac{1}{2}\dot{u}^{2}-\frac{1}{2}u^{2}\dot{v}^{2}+V_{0}\frac{\bar{m}}{4}%
u^{2}+V_{0}\frac{e^{12v}}{u^{2}}  \label{FR.20}
\end{equation}%
\begin{equation}
E=\frac{1}{2}\dot{u}^{2}-\frac{1}{2}u^{2}\dot{v}^{2}-V_{0}\frac{\bar{m}}{4}%
u^{2}-V_{0}\frac{e^{12v}}{u^{2}}.  \label{FR.21}
\end{equation}%
where $\bar{m}=-28\Lambda ~,~V_{0}=-\frac{1}{42}.$ \newline
The Euler-Lagrange equations provide the following equations of motion:%
\begin{align}
\ddot{u}+u\dot{v}^{2}-\frac{V_{0}\bar{m}}{2}u+2V_{0}\frac{e^{12v}}{u^{3}}& =0
\label{FR.22} \\
\ddot{v}+\frac{2}{u}\dot{u}\dot{v}+12V_{0}\frac{e^{12v}}{u^{4}}& =0.
\label{FR.23}
\end{align}%
The Noether integrals (\ref{NI.b5}),(\ref{NI.b6}) and the Ermakov-Lewis
invariant (\ref{NI.b6b}) in the coordinate system $\left\{ u,v\right\} $ are
\begin{align}
I_{+}& =\frac{1}{\lambda }e^{2\lambda t}E-e^{2\lambda t}u\dot{u}+\lambda
e^{2\lambda t}u^{2}  \label{FR.24} \\
I_{-}& =\frac{1}{\lambda }e^{-2\lambda t}E-e^{-2\lambda t}u\dot{u}+\lambda
e^{-2\lambda t}u^{2}.  \label{FR.25}
\end{align}%
\begin{equation}
\phi =u^{4}\dot{v}^{2}+4V_{0}e^{12v}.  \label{FR.261}
\end{equation}%
where $\lambda =\frac{1}{2}\sqrt{\frac{2}{3}\Lambda }.$

Using the Ermakov-Lewis Invariant (\ref{FR.261}), the Hamiltonian (\ref%
{FR.21}) and equation (\ref{FR.22}) are written:%
\begin{align}
\frac{1}{2}\dot{u}^{2}-V_{0}\frac{m}{8}u^{2}-\frac{1}{2}\frac{\phi }{u^{2}}&
=E  \label{FR.27} \\
\ddot{u}-\frac{V_{0}m}{4}u+\frac{\phi }{u^{3}}& =0.  \label{FR.28}
\end{align}%
The solution of (\ref{FR.28}) has been given by Pinney \cite{Pinney} and it
is the following:%
\begin{equation}
u\left( t\right) =\left( u_{1}e^{2\lambda t}+u_{2}e^{-2\lambda
t}+2u_{3}\right) ^{\frac{1}{2}}  \label{FR.29a}
\end{equation}%
where~$u_{1-3}$. From the Hamiltonian constrain (\ref{FR.27}) and the
Noether Integrals (\ref{FR.24}),(\ref{FR.25}) we find%
\begin{equation*}
E=-2\lambda u_{3}~,~I_{+}=2\lambda u_{2}~,~I_{-}=2\lambda u_{1}.
\end{equation*}%
Replacing (\ref{FR.29a}) in the Ermakov-Lewis Invariant (\ref{FR.261}) and
assuming $\phi \neq 0~$we find:%
\begin{equation}
\exp \left( v\left( t\right) \right) =2^{\frac{1}{6}}\phi ^{\frac{1}{12}%
}e^{-A\left( t\right) }\left( 4V_{0}+e^{-12A\left( t\right) }\right) ^{-%
\frac{1}{6}}  \label{FR.29}
\end{equation}%
where
\begin{equation}
A\left( t\right) =\arctan \left[ \frac{2\lambda }{\sqrt{\phi }}\left(
u_{1}e^{2\lambda t}+u_{3}\right) \right] +4\lambda ^{2}u_{1}\sqrt{\phi }.
\label{FR.31}
\end{equation}%
Then the solution is
\begin{equation}
a^{2}\left( t\right) =2^{-\frac{1}{3}}\phi ^{\frac{1}{12}}e^{-A\left(
t\right) }\left( 4V_{0}+e^{-12A\left( t\right) }\right) ^{-\frac{1}{6}%
}\left( u_{1}e^{2\lambda t}+u_{2}e^{-2\lambda t}+2u_{3}\right) ^{\frac{1}{2}}
\label{FR.32A}
\end{equation}%
where from the singularity condition $x\left( 0\right) =0~$we have the
constrain~$u_{1}+u_{2}+2u_{3}=0$ , or
\begin{equation}
2E-\left( I_{+}+I_{-}\right) =0.
\end{equation}%
At late enough time we find $A\left( t\right) \simeq A_{0}$, which implies $%
a^{2}(t)\propto e^{\lambda t}.$

In the case where $\phi =0$ equations (\ref{FR.27}),(\ref{FR.28}) describe
the hyperbolic oscillator and the solution is%
\begin{equation}
u\left( t\right) =\sinh \lambda t~,~2E=\lambda ^{2}.
\end{equation}%
From the Ermakov-Lewis Invariant we have%
\begin{equation}
\exp \left( v\left( t\right) \right) =\left( \frac{\lambda \sinh \lambda t}{%
\lambda v_{1}\sinh \lambda t-12\sqrt{\left\vert V_{0}\right\vert }%
e^{-2\lambda t}}\right) ^{\frac{1}{6}}
\end{equation}%
where $v_{1}$ is a constant. The analytical solution is%
\begin{equation}
a^{2}\left( t\right) =\left( \frac{\lambda \sinh ^{7}\lambda t}{\lambda
v_{1}\sinh \lambda t-12\sqrt{\left\vert V_{0}\right\vert }e^{-2\lambda t}}%
\right) ^{\frac{1}{6}}
\end{equation}

\section{Noether symmetries in spatially non flat $f(R)$ models}

\label{Nonf}

In this section we study further the Noether symmetries in non flat $f(R)$
cosmological models. In the context of a FRW spacetime the Lagrangian of the
overall dynamical problem and the Ricci scalar are
\begin{equation}
L=6f^{\prime }a\dot{a}^{2}+6f^{\prime \prime }\dot{R}a^{2}\dot{a}%
+a^{3}\left( f^{\prime }R-f\right) -6Kaf^{\prime }  \label{NF.01}
\end{equation}

\begin{equation}
R=6\left( \frac{\ddot{a}}{a}+\frac{\dot{a}^{2}+K}{a^{2}}\right)
\end{equation}%
where $K$ is the spatial curvature. Note that the two dimensional metric is
given by eq.(\ref{FR.03}) while the \textquotedblright
potential\textquotedblright\ in the Lagrangian takes the form
\begin{equation}
V_{K}(a,R)=-a^{3}(f^{\prime }R-f)+Kaf^{\prime }.
\end{equation}
Based on the above equations and using the theoretical formulation presented
in section \ref{LieN}, we find that the $f(R)$ models which admit non
trivial Noether symmetries are the $f(R)=(R-2\Lambda )^{3/2}$, $%
~f(R)=R^{3/2} $ and $f(R)=R^{2}$. The Noether symmetries can be found in
section \ref{LieN}.

In particular, inserting $f(R)=(R-2\Lambda )^{3/2}$ into the Lagrangian (\ref%
{NF.01})~and changing the variables from $(a,R)$ to $(x,y)$ [see section \ref%
{subs}] we find
\begin{equation}
L=\dot{x}\dot{y}+V_{0}\left( y^{3}+\bar{m}xy\right) -\bar{K}y
\end{equation}%
\begin{equation}
E=\dot{x}\dot{y}-V_{0}\left( y^{3}+{\bar{m}}xy\right) +\bar{K}y
\end{equation}%
where $\bar{K}=3(6^{1/3}K)$. Therefore, the equations of motion are
\begin{eqnarray*}
\ddot{x}-3V_{0}y^{2}-\bar{m}V_{0}x+\bar{K} &=&0 \\
\ddot{y}-\bar{m}V_{0}y &=&0\;.
\end{eqnarray*}%
The constant term $\bar{K}$ appearing into the first equation of motion is
not expected to affect the Noether symmetries (or the integrals of motion).
Indeed we find that the corresponding Noether symmetries coincide with those
of the spatially flat $f(R)=(R-2\Lambda )^{3/2}$ model. However, in the case
of $K\neq 0$ (or $\bar{K}\neq 0$) the analytical solution for the $x$%
-variable is written as
\begin{equation}
x_{K}(t)\equiv x(t)+\frac{{\bar{K}}}{\omega ^{2}}
\end{equation}%
where $x(t)$ is the solution of the flat model $K=0$ (see section \ref{subs}%
). Note that the solution of the $y$-variable remains unaltered.

Similarly, for the case of the $f(R)=R^{3/2}~$model the analytical solution
is
\begin{equation}
z_{K}\left( t\right) =z\left( t\right) +\bar{K}
\end{equation}%
where $z\left( t\right) $ is the solution of the spatially flat model (see
section \ref{subs1}).

\section{Conclusion}

\label{Conc}

In the literature the functional forms of $f(R)$ of the modified $f(R)$
gravity models are mainly defined on a phenomenological basis. In this
article we use the Noether symmetry approach to constrain these models with
the aim to utilize the existence of non-trivial Noether symmetries as a
selection criterion that can distinguish the $f(R)$ models on a more
fundamental level. Furthermore the resulting Noether integrals can be used
to provide analytic solutions.

In the context of $f(R)$ models, the system of the modified field equations
is equivalent to a two dimensional dynamical system moving in $M^{2}$ (mini
superspace) under the constraint $\bar{E}=$constant. Following the general
methodology of \cite{TsamGRG,Tsam10}, we require that the two dimensional
system admits extra Noether symmetries. This requirement fixes the $f\left(
R\right) $ function and the analytical solutions are computed. It is
interesting that two well known dynamical systems appear: the anharmonic
oscillator and the Ermakov-Pinney system. We recall that the field equations
of the $\Lambda -$cosmology is equivalent with that of the hyperbolic
oscillator.

\ack This research was partially funded by the University of Athens Special
Account of Research Grants no 10812.

\appendix

\section{Special solutions for the Power law model $R^{n}$}

\label{AppendixA}

The case $f\left( R\right) =R^{n}$ is not Liouville integrable via Noether
point symmetries. The zero order invariant will be used in order to find
special solutions. Inserting $f(R)=R^{n}~,~\left( n\neq 0,1,\frac{3}{2},%
\frac{7}{8}\right) ~$into eq.(\ref{SF.50}) we obtain%
\begin{equation}
L\left( a,\dot{a},R,\dot{R}\right) =6naR^{n-1}\dot{a}^{2}+6n\left(
n-1\right) a^{2}R^{n-2}\dot{a}\dot{R}+\left( n-1\right) a^{3}R^{n}
\end{equation}%
and the modified field equations are%
\begin{equation}
\ddot{a}+\frac{1}{a}\dot{a}^{2}-\frac{1}{6}aR=0  \label{RN.01}
\end{equation}%
\begin{equation}
\ddot{R}+\frac{n-2}{R}\dot{R}^{2}-\frac{1}{n-1}\frac{R}{a^{2}}\dot{a}^{2}+%
\frac{2}{a}\dot{a}\dot{R}-\frac{\left( n-3\right) }{6n\left( n-1\right) }%
R^{2}=0  \label{RN.02}
\end{equation}%
\begin{equation}
E=6naR^{n-1}\dot{a}^{2}+6n\left( n-1\right) a^{2}R^{n-2}\dot{a}\dot{R}%
-\left( n-1\right) a^{3}R^{n}.  \label{RN.03}
\end{equation}

The Noether symmetry (\ref{NS.07}) is also and a Lie symmetry, hence we have
the zero order invariants%
\begin{equation}
a_{0}=at^{-N}~,~R_{0}=Rt^{-2}.
\end{equation}

Applying the zero order invariants in the field equations (\ref{RN.01})-(\ref%
{RN.03}) and in the Noether integral (\ref{NI.07}) we have the following
results.

The dynamical system admit a special solution of the form
\begin{equation}
a=a_{0}t^{N}~,~R=6N\left( 2N-1\right) t^{-2}
\end{equation}%
where the constants $N,$ $E$ $\ $and $I_{7}$ are%
\begin{equation*}
N=\frac{1}{2}~,~E=0~,~I_{7}=0
\end{equation*}%
or
\begin{equation*}
N=-\frac{\left( 2n-1\right) \left( n-1\right) }{n-2}~,~E=0~,~I_{7}=0
\end{equation*}%
or%
\begin{equation*}
N=\frac{2}{3}n~,~E=\left( \frac{12n}{9}\right) ^{n}\left( 4n-3\right)
^{n-1}\left( 13n-8n^{2}-3\right) a_{0}^{3}~,~I_{7}=0.
\end{equation*}

Another special solution is the deSitter solution for $n=2$%
\begin{equation}
a=a_{0}e^{H_{0}t}~,~R=12H_{0}^{2}
\end{equation}%
where $I_{7}=0~$\ and the spacetime is empty i.e. $E=0$.

\end{document}